\documentclass[8.5pt,twoside,twocolumn]{article}
\usepackage[super,sort&compress,comma]{natbib}
\usepackage{mhchem}
\usepackage{times,mathptm}
\usepackage{sectsty}
\usepackage{balance}

\usepackage{graphicx} 
\usepackage{lastpage}
\usepackage[format=plain,justification=raggedright,singlelinecheck=false,font=small,labelfont=bf,labelsep=space]{caption}
\usepackage{fancyhdr}
\begin{document}

\thispagestyle{plain} \fancypagestyle{plain}{
\renewcommand{\headrulewidth}{1pt}}
\renewcommand{\thefootnote}{\fnsymbol{footnote}}
\renewcommand\footnoterule{\vspace*{1pt}%
\hrule width 3.4in height 0.4pt \vspace*{5pt}}
\setcounter{secnumdepth}{5}

\makeatletter
\def\subsubsection{\@startsection{subsubsection}{3}{10pt}{-1.25ex plus -1ex minus -.1ex}{0ex plus 0ex}{\normalsize\bf}}
\def\paragraph{\@startsection{paragraph}{4}{10pt}{-1.25ex plus -1ex minus -.1ex}{0ex plus 0ex}{\normalsize\textit}}
\renewcommand\@biblabel[1]{#1}
\renewcommand\@makefntext[1]%
{\noindent\makebox[0pt][r]{\@thefnmark\,}#1} \makeatother
\renewcommand{\figurename}{\small{Fig.}~}
\sectionfont{\large}
\subsectionfont{\normalsize}

\fancyfoot{}
\fancyfoot[RO]{\footnotesize{\sffamily{1--\pageref{LastPage}
~\textbar  \hspace{2pt}\thepage}}}
\fancyfoot[LE]{\footnotesize{\sffamily{\thepage~\textbar\hspace{3.45cm}
1--\pageref{LastPage}}}} \fancyhead{}
\renewcommand{\headrulewidth}{1pt}
\renewcommand{\footrulewidth}{1pt}
\setlength{\arrayrulewidth}{1pt} \setlength{\columnsep}{6.5mm}
\setlength\bibsep{1pt}

\twocolumn[
  \begin{@twocolumnfalse}
\noindent\LARGE{\textbf{Titanium-capped carbon chains as promising
new hydrogen storage media}} \vspace{0.6cm}

\noindent\large{\textbf{Chun-Sheng Liu,\textit{$^{ab}$} Hui
An,\textit{$^{ab}$} and Zhi
Zeng$^{\ast}$\textit{$^{ab}$}}}\vspace{0.5cm}

\noindent\textit{\small{\textbf{Received Xth XXXXXXXXXX 20XX,
Accepted Xth XXXXXXXXX 20XX\newline First published on the web Xth
XXXXXXXXXX 200X}}}

\noindent \textbf{\small{DOI: 10.1039/b000000x}} \vspace{0.6cm}

\noindent \normalsize{The capacity of Ti-capped \emph{sp} carbon
atomic chains for use as hydrogen storage media is studied using
first-principles density functional theory. The Ti atom is strongly
attached at one end of the carbon chains via \emph{d}-\emph{p}
hybridization, forming stable TiC$_\emph{n}$ complexes. We
demonstrate that the number of adsorbed H$_2$ on Ti through Kubas
interaction depends upon the chain types. For polyyne (\emph{n}
even) or cumulene (\emph{n} odd) structures, each Ti atom can hold
up to five or six H$_2$ molecules, respectively. Furthermore, the
TiC$_5$ chain effectively terminated on a C$_{20}$ fullerene can
store hydrogen with optimal binding of 0.52 eV/H$_2$. Our results
reveal a possible way to explore high-capacity hydrogen storage
materials in truly one-dimensional carbon structures.}
\vspace{0.5cm}
 \end{@twocolumnfalse}
  ]

\section{Introduction}


\footnotetext{\textit{$^{a}$~AKey Laboratory of Materials Physics,
Institute of Solid State Physics, Chinese Academy of Sciences, Hefei
230031, People's Republic of China. Fax: +86-551-5591434; Tel:
+86-551-5591407; E-mail: zzeng@theory.issp.ac.cn}}
\footnotetext{\textit{$^{b}$~Graduate School of the Chinese Academy
of Sciences, Beijing 100094, People's Republic of China. }}

To develop economical hydrogen energy, carbon nanostructures with
\emph{sp}$^2$-like bonding functionalized by transition metal
(TM),\cite{ref01,ref02,ref03,ref04,ref05,ref06} alkali metal (AM)
\cite{ref07,ref08} and alkaline-earth metal (AEM) atoms \cite{ref09}
have been expected to be promising storage materials due to their
light weights and large surface areas. Recently, Iijima \emph{et
al.} \cite{ref10} transformed graphene to single
\emph{sp}-hybridized carbon chains containing 16 carbon atoms. In
addition, chains with metal atoms connected to the ends have been
previously generated,\cite{ref11} and their magnetic, electronic,
and transport properties have been studied
extensively.\cite{ref12,ref13} These results inspired us to consider
whether metal-capped carbon atomic chains are efficient hydrogen
storage media. In this study, we conduct theoretical studies of
high-capacity hydrogen storage media consisting of a Ti atom capped
on two kinds of atomic carbon chains, cumulene (with double C-C
bonds) or polyyne (with alternating singlet and triplet bonds). The
number of adsorbed H$_2$ molecules is only determined by the type of
chain. Each Ti atom in TiC$_5$ and TiC$_8$ can bind up to six and
five H$_2$ molecules, respectively, corresponding to storage
capacities of 10 wt \% and 6.5 wt \%. The average binding energy of
H$_2$ molecules on TiC$_5$ (TiC$_8$) is 0.59 eV/H$_2$ (0.57
eV/H$_2$), which is between the physisorption and chemisorption
energies.
\section{Computational method}

Numerical calculations were performed using spin-polarized density
functional theory (DFT) with the Perdew-Wang (1991)
exchange-correlation function,\cite{ref14} as implemented in the
DMol$^3$ package (Accelrys Inc.).\cite{ref15} A double numerical
atomic orbital augmented by \emph{d}-polarization functions (DNP)
was employed as the basis set. In the self-consistent field
calculations, the electronic-density convergence threshold was set
to 1$\times$10$^{-6}$ electron/\AA$^3$. Geometric optimization was
performed with convergence thresholds of 10$^{-5}$ Ha for the
energy, 2$\times$10$^{-3}$ Ha/\AA $ $ for the force, and 10$^{-4}$
\AA $ $ for the atomic displacements. We performed normal-mode
analysis of the obtained structures to ensure that the structures
optimized without any symmetry constraints were true minima of the
potential-energy surface.

\section{Results and discussion}

We first consider the bonding energetics of a single Ti atom at one
end of the C$_n$ (\emph{n}=5-16) chains. The binding energies of Ti,
shown in Fig. 1(a), are much larger than those of Ti on
\emph{sp}$^2$- hybridized carbon
materials,\cite{ref02,ref03,ref04,ref05} and the adsorption
strengths are thus strong enough to maintain stable Ti-C$_\emph{n}$
structures at room temperature. Moreover, the Ti binding energies
exhibit an interesting odd-even oscillatory behavior, consistent
with the trend of Mulliken charge [see Fig. 1(a)]. This behavior
suggests that the bonding between the Ti atom and carbon chains has
a significant ionic characteristic. In addition, the configurations
of TiC$_\emph{n}$ complexes exhibit significant differences for odd
or even values of \emph{n}. For instance, the C-C distances in
TiC$_5$, depicted in Fig .1(b), are rather uniform. In contrast, the
bond lengths of TiC$_8$ alternate, consistent with $\sigma$-$\pi$
bonding [Fig. 1(c)]. Contour plots of the total electronic charge
density confirm these bonding patterns.

\begin{figure}[h]
\centering
  \includegraphics[height=8cm]{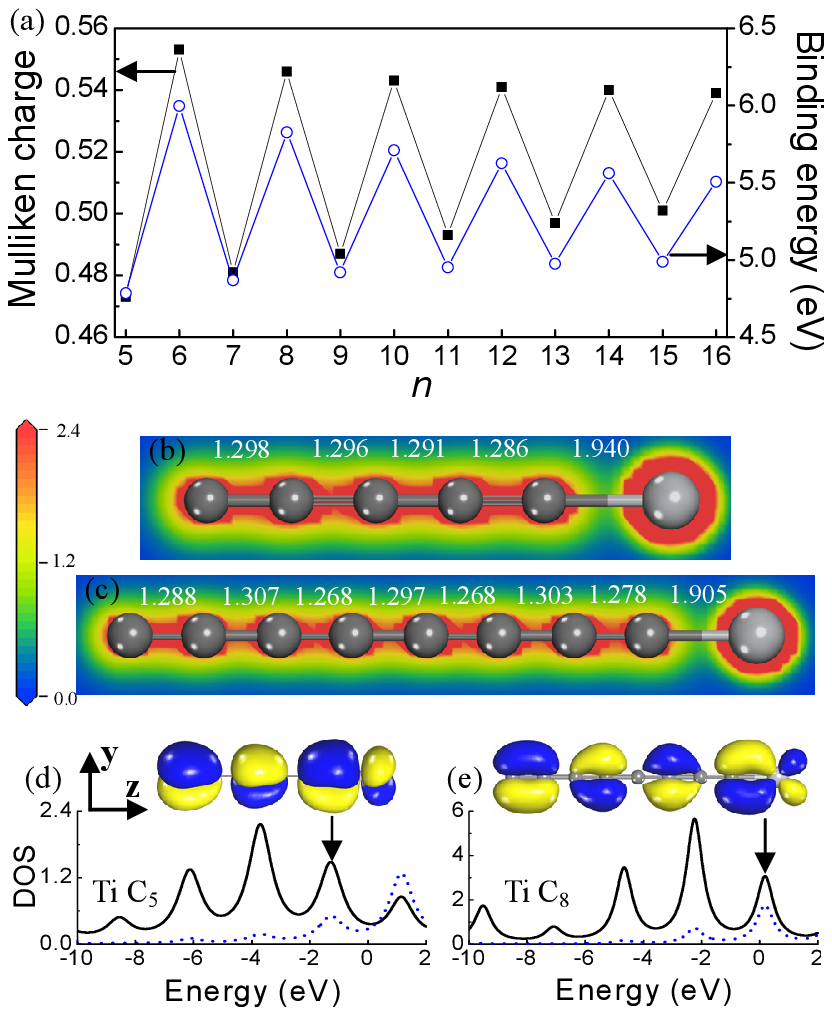}
  \caption{(a) Mulliken charge and binding energy of TiC$_n$ as a
function of the number of carbon atoms. The optimized bond lengths
(in \AA) and the contour plot for the total electronic charge
density of (b) TiC$_5$ and (c) TiC$_8$. The large and small balls
represent the Ti and C atoms, respectively. Calculated partial
density of states (PDOS) of (d) TiC$_5$ and (e) TiC$_8$ in units of
state/eV. The solid black lines and dotted blue lines represent the
C-2\emph{p} and Ti-3\emph{d} states, respectively. Insets of (d) and
(e) show the bonding orbitals (the two colors denote $\pm$ signs of
the wave function) corresponding to states indicated by black
arrows.}
\end{figure}

To elucidate the nature of the bonding, the projected density of
states (DOS) shown in Fig. 1(d) illustrates the bonding orbitals of
the C$_5$ (C$_8$) and Ti atom result from the hybridization of the
carbon chain $\pi$ orbitals and the Ti-\emph{d} orbitals. The
C-\emph{p} and Ti-\emph{d} hybridization is evidently stronger in
TiC$_8$ than in TiC$_5$, in accordance with the larger binding
energy of TiC$_8$. Since odd-numbered and even-numbered free chains
have full and half-occupied HOMOs (highest occupied molecular
orbitals), respectively, different states are responsible for the
bonding of the Ti atom to C$_5$ or C$_8$. From the isosurface plots
of these states [Fig. 1(d)], it is clearly that the LUMO (a linear
combination of degenerate $\pi$$_{p_x}$$^\ast$ and
$\pi$$_{p_y}$$^\ast$ antibonding orbitals) of C$_5$ hybridizes with
two Ti-\emph{d} orbitals to create a new $\pi$-symmetry bond.
However, for TiC$_8$ there is an overlap between one Ti-\emph{d}
orbital and the $\pi$$_{p_y}$$^\ast$ antibonding state of C$_8$. As
mentioned above, the bonding mechanisms for TiC$_5$ and TiC$_8$ seem
to be very similar to the Dewar coordination.\cite{ref16}

The stability of the TiC$_5$ and TiC$_8$ complexes was further
tested by normal mode analysis, and no negative mode was found.
However, it remains to be shown whether it is possible to realize
linear C$_\emph{n}$-Ti chains experimentally. To address this issue,
we calculated the minimum-energy path (MEP) for different reaction
processes with the nudged elastic band method to determine the
energy barrier.\cite{ref17} The image number considered is 16 to
ensure that the obtained MEP is correct. In Figs. 2(a) and 2(b), the
MEPs show that a single Ti atom is attached to the right end of
C$_5$ and C$_8$ without any energy barrier. The other end is free
and ready to be attached to any other structure with a high
surface-to-volume ratio. The main Ti vibrational frequencies, 623
cm$^{-1}$ in TiC$_8$ and 359 cm$^{-1}$ in TiC$_5$, correspond to the
stretching modes of C-Ti. These characteristic modes will provide a
reference for the Raman/IR spectra of the synthesized materials. To
this end, a monatomic carbon chain, realized by removing the carbon
atoms from the graphene through energetic electron irradiation
inside a transmission electron microscope,\cite{ref10} can be used
as the initial material in the fabrication process.
\begin{figure}[h]
\centering
  \includegraphics[height=3cm]{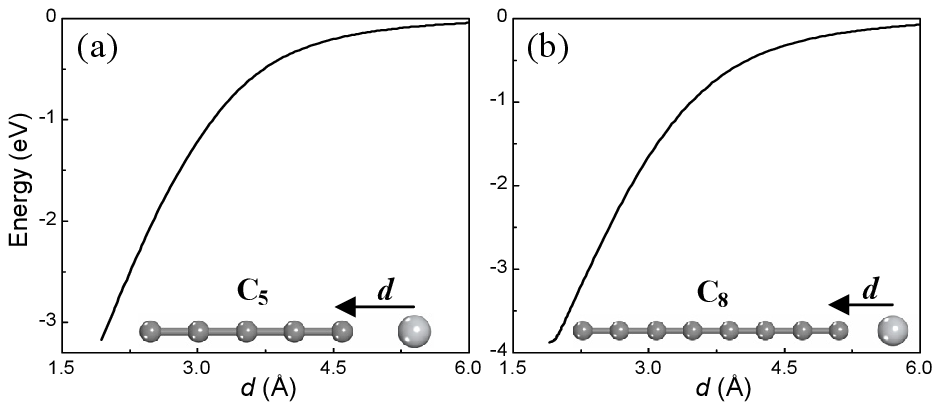}
  \caption{Variation in the total-energy difference of a Ti atom
attached at a distance \emph{d} from the right free end of the
carbon chain.}
\end{figure}

Now we investigate the interaction between these complexes and
hydrogen molecules. The first H$_2$ molecule is found to dissociate
on both TiC$_5$ and TiC$_8$ complexes and bind atomically to Ti,
with interatomic H distances of 3.003 \AA$ $ [Fig. 3(a)] and 2.748
\AA$ $ [Fig. 3(b)], respectively. Such a dissociation is observed in
most TM-hydrogen binding [Ref. 2] due to excessive charge transfer
from the TM to the antibonding $\sigma$$^\ast$ state of H$_2$.
Additional H$_2$ molecules, however, bind molecularly around the Ti
atom, since the charge transfer per H$_2$ molecule is not enough to
destabilize the dihydrogen state when more hydrogen molecules are
added to the system.

For TiC$_5$, several initial configurations of the second H$_2$
molecule were considered in the search for the lowest energy
structure. The structure in which one of the H$_2$ molecules binds
as a dihydride is 0.09 eV higher in energy than the structure in
which both H$_2$ molecules bind in molecular form [Fig. 3(a)]. We
continue to introduce successive hydrogen molecules near the Ti
atom. Surprisingly, it is energetically favorable for the TiC$_5$
complex to adsorb 6 H$_2$ molecules, corresponding to a $\sim$10 wt
\% gravimetric density. A close examination of the geometry of the
TiC$_5$(H$_2$)$_6$ configuration shows that the adsorbed hydrogen
molecules are divided into two sets [Fig. 2(a)]. First, the molecule
on the top binding site, which is vertical to the carbon chain axis,
has a bond length of 0.808 \AA$ $ and the largest bonding energy,
0.82 eV. Second, the H-H bonds of the side H$_2$ molecules lie
almost parallel to the chain axis, with the lower H atoms of each
H$_2$ molecule tending to tilt toward the chain. The bond lengths
between the Ti and the lower (upper) H atoms are 1.910 \AA$ $ (1.970
\AA). The side molecules are more weakly bound, with a binding
energy of 0.54 eV/H$_2$ and an elongated H-H bond distance of 0.810
\AA$ $, on average. All of these features suggest that different
bonding mechanisms (to be discussed below) for the two sets of H$_2$
molecules on TiC$_5$.

On the other hand, when the second H$_2$ molecule is bound to
TiC$_8$, the structural configuration shown in Fig. 3(b) is more
energetically favorable than its isomer (not shown) by 1.02 eV. As
we add a third H$_2$ molecule close to the Ti atom, all of the
hydrogen molecules are almost parallel to the chain and are bonded
molecularly. Interestingly, the H-H bonds turn nearly perpendicular
to the chain when the fourth H$_2$ molecule is attached. The maximum
number of H$_2$ molecules that could be bound to the TiC$_8$ complex
is 5, which corresponds to a hydrogen wt \% of 6.5. Note that the
adsorption of the fifth H$_2$ molecule does not affect the degree of
the bond-length elongation of the side hydrogen molecules [Fig.
3(b)]. The H-H bond of the top H$_2$ molecule is 0.756 \AA$ $ and
the bond length between this H$_2$ molecule and the Ti atom is 2.334
\AA, leading to an adsorption energy of 0.29 eV, which is only about
half the value of the side molecules (0.66 eV).

\begin{figure}[h]
\centering
  \includegraphics[height=8.5cm]{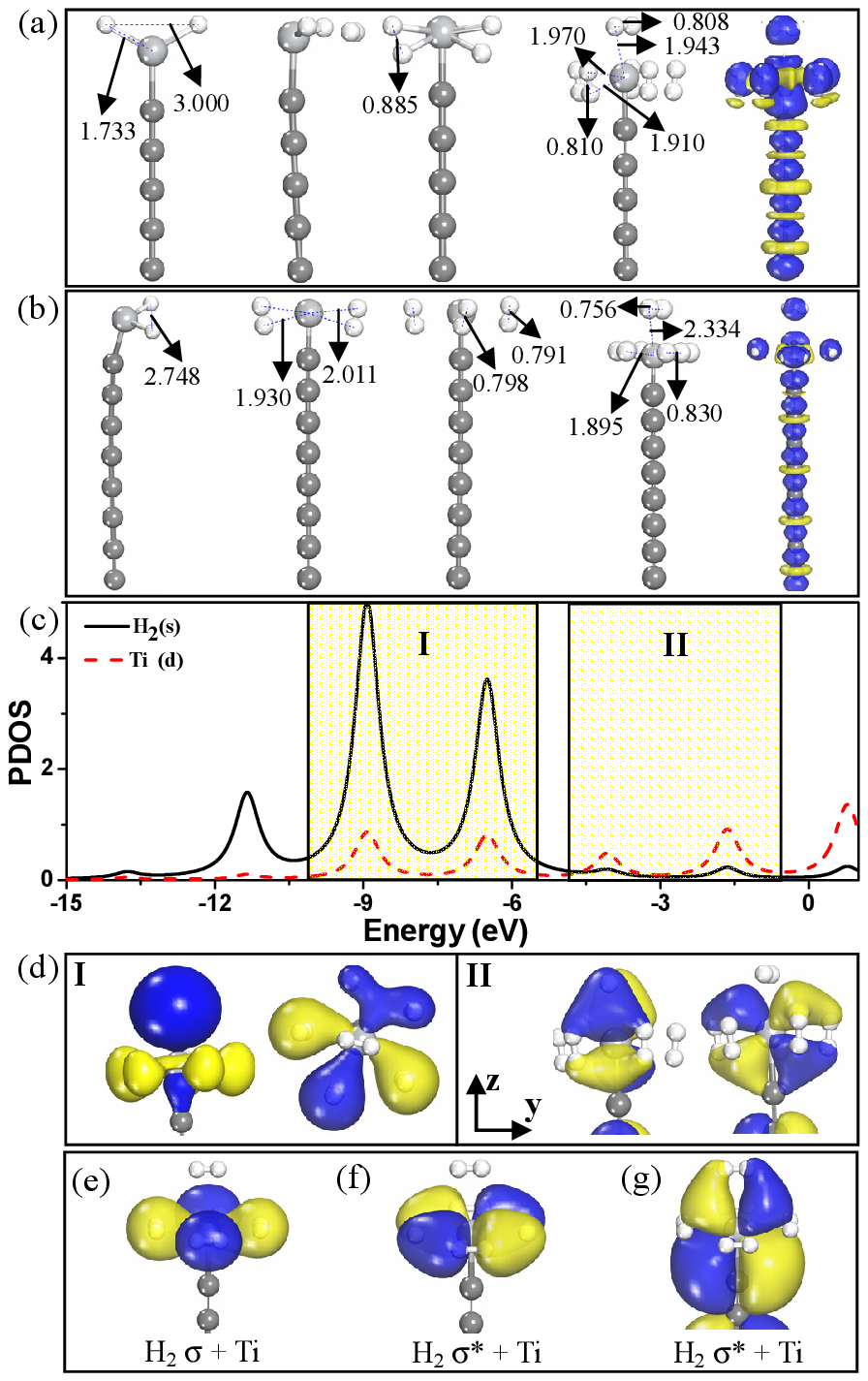}
  \caption{Optimized configurations of H$_2$ molecules adsorbed to
the Ti atom on (a) C$_5$ and (b) C$_8$ along with the typical bond
lengths (in \AA). The large, medium and small balls represent the
Ti, C and H atoms, respectively. The right panels of (a) and (b)
show the deformation electron densities (molecular charge densities
minus atomic charge densities) for the TiC$_5$(H$_2$)$_6$ and
TiC$_8$(H$_2$)$_5$ complexes. The deformed density marked in blue
corresponds to the region that contains excess electrons, while that
marked in yellow indicates electron loss. The isovalue equals 0.1
\emph{e}/\AA$^3$. (c) PDOS for six hydrogen molecules attached to
the Ti atom on C$_5$. (d) Isosurfaces of major molecular orbitals
corresponding to states in region I and II of panel (c). Panels (e)
and (f) show the $\sigma$ and $\sigma$$^\ast$ orbitals,
respectively, for the side hydrogen molecules hybridized with the
Ti-\emph{d} orbitals in TiC$_8$(H$_2$)$_5$. (g) The bonding orbital
for the top hydrogen molecule.}
\end{figure}

Insight into the nature of Ti-H$_2$ bonding and the orientation of
H$_2$ molecules can be gained from the DOS and molecular orbital
(MO) analyses. The metal-dihydrogen binding of TiC$_5$(H$_2$)$_6$,
shown in Fig. 3(c), stems primarily from the Kubas
interaction.\cite{ref18} In the energy range from -10 to -6 eV
(region I), the \emph{d} orbitals of Ti are hybridized with the
$\sigma$ orbitals of the hydrogen molecules, resulting in charge
transfer from the hydrogen to the metal. Region II of Fig. 3(c)
highlights that the hybridization of the Ti-\emph{d} orbitals with
the H$_2$ $\sigma$$^\ast$-antibonding orbitals is responsible for
keeping the side H$_2$ molecules parallel and the top H$_2$ molecule
perpendicular to the TiC$_5$ chain. The isosurface plots of these
states, as shown in Fig. 3(d), clearly confirm that the bonding is
dominated by the overlap between the Ti-\emph{d} and H$_2$-\emph{s}
states. To better understand the orientation of the side H$_2$
molecules, we also plotted the deformation charge in Fig. 3(a). The
charge distribution around the lower H sites, marked in yellow,
indicates that these H atoms lose a greater amount of charge. With
the help of Mulliken charge population analysis, we identified that
the upper H atom has a charge of 0.04 \emph{e}, while the lower one
carries 0.12 \emph{e} due to charge transfer from H to the C atom.
Thus, the distances between the Ti and the upper or lower H depend
on the strength of the Coulomb interaction between the H and C
atoms. The Kubas interaction is likewise found in
TiC$_8$(H$_2$)$_5$. Figure 3(e) shows that the Ti-\emph{d} orbitals
interact with the $\sigma$ orbitals of the H$_2$ molecules. The
hybridization between the Ti-\emph{d} orbitals and the side H$_2$
molecules' $\sigma$$^\ast$-antibonding orbitals plays a role in
keeping them planar [Fig. 3(f)], in contrast to the situation in
TiC$_5$(H$_2$)$_6$. The top hydrogen in TiC$_8$(H$_2$)$_5$ interacts
with Ti rather weakly compared to the rest of the hydrogen
molecules, as the relevant MO is more heavily polarized toward the
C$_8$ than toward the H$_2$ [see Fig. 3(g)]. The Kubas interaction
mentioned above leads to variations in the effective charge of the
Ti atom. The Mulliken charge of the Ti atom varies from positive to
negative as the number of hydrogen molecules increases. When the
fifth (sixth) H$_2$ is adsorbed onto TiC$_8$ (TiC$_5$), the Ti atom
carries -0.13 (-0.49) electrons.

The above results demonstrate that the interaction between one Ti
atom and the carbon chain is very strong, and the binding energy of
H$_2$ on TiC$_n$ is optimal for room temperature application. Now
the question arises: can the storage performance of TiC$_n$ be
influenced by introducing another Ti? We first discuss the second Ti
atom saturating the free end of TiC$_n$ (\emph{n}= 5, 8). The
binding energy for Ti to the TiC$_5$ (TiC$_8$) is 4.31 eV (4.39 eV),
which is obviously much large than that of Ti on the \emph{sp}$^2$
carbon nanostructures.\cite{ref02,ref03,ref04,ref05} However,
further functionalization through Ti atoms bridging other C-C bonds
makes the chain unstable, due to the fact that the small separation
between Ti atoms leads to stronger Ti-Ti coupling. At the end, the
linear chains are transformed to the clusters composed of C and Ti
atoms, which is also confirmed by other theoretical
studies.\cite{ref19} We therefore only consider two Ti atoms each
capping one end of the C$_n$.

Next, we studied the H$_2$ storage capacity of the TiC$_n$Ti
complex. The TiC$_5$Ti and TiC$_8$Ti can adsorb 12 and 10 H$_2$
molecules, respectively, with average binding energies of 0.53 and
0.55 eV/H$_2$. As shown in Figs. 4(a) and 4(b), the gravimetric
densities of stored H$_2$ molecules for TiC$_n$Ti becomes 13.3 and
9.4 wt \% when \emph{n}=5 and \emph{n}=8, respectively. Analysis of
the electronic structures of the TiC$_5$Ti-(H$_2$)$_{12}$ and
TiC$_8$Ti-(H$_2$)$_{10}$ shows that the Kubas interaction between
H$_2$ and TiC$_n$ mentioned above also applies to the TiC$_n$Ti
complex. Note that introduction of the second Ti atom to TiC$_n$ not
only keeps the number of adsorbed H$_2$ per Ti on odd- or
even-numbered chains, but also remains the binding energies of the
H$_2$ with metal atoms .
\begin{figure}[h]
\centering
  \includegraphics[height=5cm]{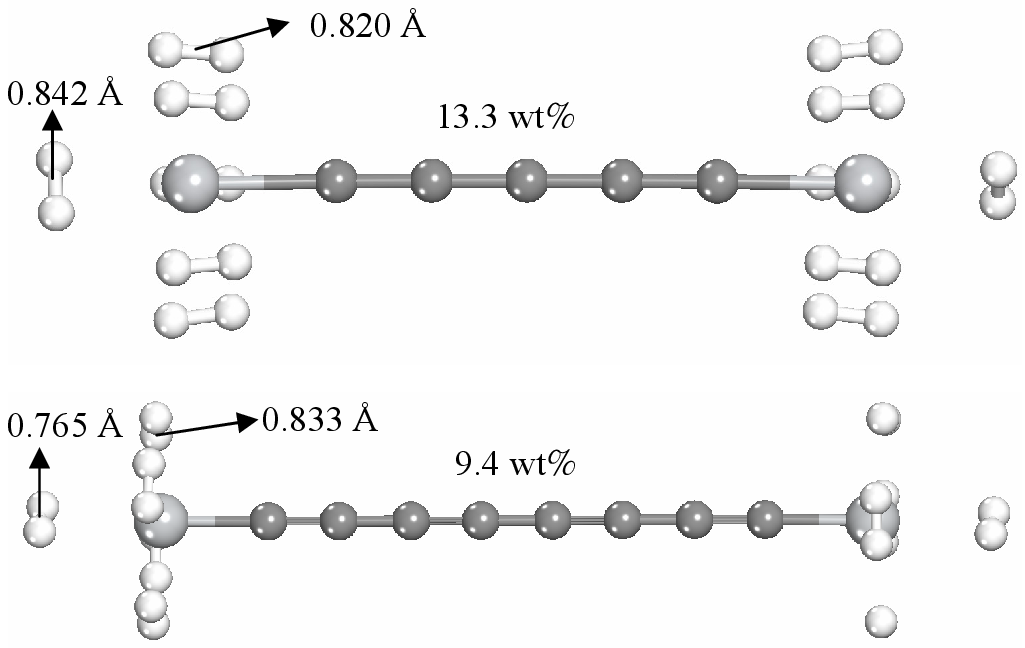}
  \caption{Geometries of (a) TiC$_5$Ti-(H$_2$)$_{12}$ and (b)
TiC$_8$Ti-(H$_2$)$_{10}$. The large, medium and small balls
represent the Ti, C and H atoms, respectively. }
\end{figure}

Because controlling the number of atoms per chain in the synthesis
of carbon chains may be difficult with today's technology, it is
important to know whether the results reported above for TiC$_5$ and
TiC$_8$ are applicable to other Ti-capped carbon chains and how they
vary with chain length. To this end, we have also studied the
potential of the TiC$_{15}$ and the TiC$_{16}$ complexes as storage
media, and we found that the number of adsorbed H$_2$ molecules is
six and five, respectively. The corresponding bond lengths of H$_2$
and Ti-H$_2$, as depicted in Fig. 5(a) and 5(b), are very similar to
those in TiC$_5$(H$_2$)$_6$ and TiC$_8$(H$_2$)$_5$. Furthermore, the
average binding energies of the H$_2$ molecules to TiC$_{15}$ and
TiC$_{16}$ are, respectively, 0.54 and 0.59 eV/H$_2$, which are very
close to the values found in the case of TiC$_5$(H$_2$)$_6$ and
TiC$_8$(H$_2$)$_5$. The bonding mechanisms between Ti and H$_2$
presented above also hold for these chains. This suggests that the
length of the chains does not affect the hydrogen adsorption
performance. In summary, the finding that \emph{a single Ti
atom-capped, odd-numbered (even-numbered) carbon chain can bind up
to six (five) hydrogen molecules} is very general.
\begin{figure}[h]
\centering
  \includegraphics[height=6cm]{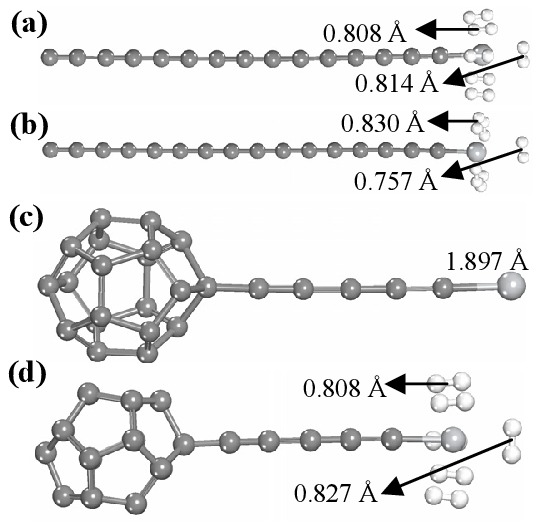}
  \caption{Geometries of (a) TiC$_{15}$(H$_2$)$_6$, (b)
TiC$_{16}$(H$_2$)$_5$, and (c) a TiC$_5$ chain terminated on the
C$_{20}$ fullerene. (d) Optimized geometry of 6 H$_2$ molecules
adsorbed on C$_{20}$(TiC$_5$). The large, medium and small balls
represent the Ti, C and H atoms, respectively. }
\end{figure}

While the above results are promising for isolated TiC$_\emph{n}$
systems, one can imagine terminating the other end of these
complexes with suitable graphitic nano-objects. These new structures
not only represent the typical interface in realistic nanostructures
produced by cluster beam deposition.\cite{ref20} Here we choose
C$_{21}$ fullerene as the end-capping candidate, as it is the most
frequently experimentally synthesized curved \emph{sp}$^2$ system.
An increasing amount of experimental \cite{ref21} evidence shows
that these hybrid \emph{sp}+\emph{sp}$^2$ carbon-based systems (with
linear chains connecting \emph{sp}$^2$-type fragments) exhibit
unusual electronic and optical properties. Figure 5(c) illustrates
that a TiC$_5$ chain can be effectively stabilized by termination on
C$_{20}$. The calculated binding energy (with respect to the
isolated TiC$_5$ plus the fully relaxed C$_{20}$) is 2.79 eV. As
shown in Fig. 5(d), the TiC$_5$ assembled on the C$_{20}$ structure
can hold 6 H$_2$ molecules, with an average binding energy of 0.52
eV/H$_2$.

\section{Conclusion}

In conclusion, using all-electron DFT calculations, we have shown
that each Ti atom adsorbed on even or odd-numbered carbon atomic
chains can bind up to five or six hydrogen molecules, respectively.
Note that the number of adsorbed H$_2$ molecules depends only on the
type of chain. We propose that the TiC$_5$ chain terminated
effectively on a C$_{20}$ fullerene can also store 6 H$_2$ molecules
with an average binding energy of 0.52 eV/H$_2$. Recent experiments
have produced junctions between a single carbon chain and two
fullerenes,\cite{ref22} which has provided a way to synthesis the
\emph{sp}+\emph{sp}$^2$ system we proposed here. We hope that the
theoretical results presented here will provide a useful reference
for the design of high-capacity hydrogen storage materials in the
laboratory.

\section{Acknowledgements}

This work was supported by the National Science Foundation of China
under Grant no. 10774148, the special funds for the Major State
Basic Research Project of China (973) under Grant no. 2007CB925004,
and the Knowledge Innovation Program of the Chinese Academy of
Sciences. Part of the calculations were performed at the Center for
Computational Science of CASHIPS.

\end{document}